\documentclass{article}

\usepackage{epsfig} 
\usepackage{amsmath}
\usepackage{amsfonts}
\usepackage{graphicx}

 \usepackage{amssymb}

\date{\today}

\newcommand{\insertplot}[5]{\begin{figure}
 \hfill\hbox to 0.05in{\vbox to #5in{\vfill
 \inputplot{#1}{#4}{#5}}\hfill}
 \hfill\vspace{-.1in}
 \caption{#2}\label{#3}
 \end{figure}}
 \newcommand{\inputplot}[3]{
 \special{ps: plotfile #1}
}
\newcounter{fig}

\renewcommand{\t}{\theta}

\newcommand{\f}{\phi}

\newcommand{\ee}{\end{equation}}
\newcommand{\eea}{\end{eqnarray}}
\newcommand{\be}{\begin{equation}}
\newcommand{\bea}{\begin{eqnarray}}

\begin{document}

\title{ 
Spinning scalar solitons in anti-de Sitter spacetime
} 
  
\author{
{\large Eugen Radu} 
and {\large Bintoro Subagyo}  \\ 
 {\small  Institut f\"ur Physik, Universit\"at Oldenburg, Postfach 2503
D-26111 Oldenburg, Germany} }

\maketitle

\begin{abstract} 

We present spinning $Q$-balls and boson stars
in four dimensional anti-de Sitter spacetime.
These are smooth, horizonless solutions for gravity coupled to 
a massive complex scalar field with a  harmonic
dependence on time and the azimuthal angle.
Similar to the flat spacetime configurations, the angular momentum is quantized.
We find that a class of solutions with a self-interaction potential
has a limit corresponding to static solitons with
axial symmetry only.
An exact solution describing spherically symmetric 
$Q-$balls in a fixed AdS background is also discussed.
\end{abstract}

\section{Introduction}
Anti-de Sitter (AdS) gravity coupled to scalar fields only, has received recently considerable 
attention.
This simple model admits both soliton 
\cite{Hertog:2004rz}, 
\cite{Hertog:2004ns}, 
\cite{Radu:2005bp},
\cite{Faulkner:2010fh},
and black hole solutions (see $e.g.$
\cite{Torii:2001pg},
\cite{Sudarsky:2002mk},
\cite{Winstanley:2002jt},
\cite{Hertog:2004dr},
\cite{Martinez:2004nb},
\cite{Park:2008zzb})
with interesting properties, the resulting picture being in strong contrast
with the one found 
in the absence of a cosmological constant.
Moreover, since scalar fields generically enter the gauge supergravity
models, the study of such solutions is also relevant
to the AdS/CFT conjecture.
 
However, most of the studies in the literature assume that the scalar fields 
are real and have the same symmetries as the underlying spacetime. 
It is interesting to ask if the known solutions in \cite{Hertog:2004rz}-\cite{Martinez:2004nb} 
can be generalized 
for a  complex scalar field. 
In formulating a  scalar field ansatz in this case, it is natural to take 
a separation of variables 
\begin{equation}
\label{Phi}
\Phi(\vec x,t)=e^{-i  \omega t}\phi(\vec x),
\end{equation}
(with $t,\vec x$ the time and space coordinates, respectively) 
such that the energy-momentum tensor is time independent
(note that in general, $\phi(\vec x)$ is a complex function as well).
This model  possesses a conserved current 
\begin{eqnarray}
j^{\mu}= - i \left( \Phi^* \partial^{\mu} \Phi 
 - \Phi \partial^{\mu}\Phi ^* \right)  ,~~~
j^{\mu} _{\ ; \, \mu}  =  0 \ ,
\end{eqnarray}
and a conserved Noether charge $Q$, which is the integral over a $t=const.$ hypersurface
of the $j^t$ component of the current.
 
Scalar field configurations carrying a global U(1) Noether charge have been extensively studied in the literature, 
for a Minkowski spacetime background and four spacetime dimensions.
While black hole solutions are rather difficult to find in this case 
\cite{Pena:1997cy},
\cite{Kleihaus:2010ep},
the spectrum of  smooth horizonless solutions with harmonic 
 time dependence is very rich.
For example, one finds (non-topological) soliton solutions even
in the absence of gravity--the so-called $Q-$balls 
\cite{Friedberg:1976me},
\cite{Coleman:1985ki}, 
\cite{Lee:1991ax}. 
There, the scalar field possesses a potential which
is necessarily non-renormalizable\footnote{It is interesting to note that $Q-$ball solutions appear in supersymmetric 
generalizations of the standard model 
\cite{Kusenko:1997zq}.
They may be responsible for the generation of
baryon number or may be regarded as candidates for dark matter  
\cite{Kusenko:1997si}. 
}.
When  gravity is coupled to $Q-$balls, boson stars arise 
(see the review work \cite{Jetzer:1991jr}).
Moreover,  in this case, solutions with rather similar properties are found even for a
potential consisting in a mass term only
\cite{Kaup:1968zz}, 
\cite{Ruffini:1969qy}, 
\cite{Mielke:1980sa}.
The self-gravity of such objects is balanced by the dispersive effect
due to the wave character of the complex scalar field.

 The study of boson stars and Q-balls in 
 AdS spacetime has received relatively little attention, only spherically symmetric solutions
 being discussed so far 
 \cite{Astefanesei:2003qy}, 
 \cite{Prikas:2004yw}, 
 \cite{Hartmann:2012wa} 
 (see, however, the planar solitons with a complex scalar field in \cite{Horowitz:2010jq}).
 While the results in \cite{Astefanesei:2003qy} for a gravitating 
 massive scalar field without self-interaction are rather similar to those valid in 
 the asymptotically flat limit, 
 the recent study 
 \cite{Hartmann:2012wa}
  has shown the existence of some new features
 in the case of Q-ball solutions.

For a vanishing cosmological constant, $\Lambda=0$, the scalar solitons admit also 
rotating generalizations 
\cite{Schunck:1996wa}, 
\cite{Yoshida:1997qf}, 
\cite{Volkov:2002aj},
\cite{Kleihaus:2005me}, 
\cite{Kleihaus:2007vk}.
These are stationary localized solutions possessing a finite mass and angular momentum.
Interestingly, their angular momentum is quantized
$J=n Q$ (with $n$ an integer), 
and the energy density exhibits a toroidal distribution.
 
The main purpose of this work is to investigate the existence of spinning scalar solitons 
for the case of a four dimensional AdS background\footnote{Rotating AdS boson stars were found 
however in $d=3$ \cite{Astefanesei:2003rw} and also in $d=5$ \cite{Dias:2011at}
dimensions, where a special ansatz proposed in \cite{Hartmann:2010pm} allows to deal with ODEs.}.
These solutions are found by solving numerically
a set of partial differential equations with suitable boundary conditions. 
Irrespective of the scalar field potential, they exhibit  the same quantization of the angular momentum
as for $\Lambda=0$.
We find that the spinning 
  solutions emerge as perturbations of the AdS spacetime
for a critical value of the frequency  which is fixed by the scalar field mass and the cosmological constant.
In the absence of a scalar field self-interaction, the basic
properties of the solutions
are rather similar to those of the asymptotically flat counterparts.
New features are found once we allow for self-interaction terms in the potential
leading to a violation of the positive energy condition.
For example, we find a class of solutions with a smooth $\omega=0$
limit, describing static axially symmetric solitons.

\section{The  model}
\subsection{The action and field equations}
 We consider the action of a self-interacting complex scalar field 
$\Phi$ coupled to Einstein gravity with a negative cosmological constant
$\Lambda=-3/\ell^2$,
\begin{equation}
\label{action}
S=\int  d^4x \sqrt{-g}\left[ \frac{1}{16\pi G}(R-2 \Lambda)
   -\frac{1}{2} g^{\mu\nu}\left( \Phi_{, \, \mu}^* \Phi_{, \, \nu} + \Phi _
{, \, \nu}^* \Phi _{, \, \mu} \right) - U( \left| \Phi \right|) 
 \right] , 
\end{equation}
where $R$ is the curvature scalar,
$G$ is Newton's constant,
the asterisk denotes complex conjugation,
and $U$ denotes the scalar field potential.

Variation of the action with respect to the metric
leads to the Einstein equations
\begin{equation}
\label{Einstein-eqs}
E_{\mu\nu}= R_{\mu\nu}-\frac{1}{2}g_{\mu\nu}R+\Lambda g_{\mu\nu} - 8 \pi G T_{\mu\nu}=0\ , 
\end{equation}  
where $T_{\mu\nu}$ is the
 stress-energy tensor  of the scalar field
\begin{eqnarray}
\label{tmunu} 
T_{\mu \nu}  
&=&
\left(
 \Phi_{, \, \mu}^*\Phi_{, \, \nu}
+\Phi_{, \, \nu}^*\Phi_{, \, \mu} 
\right )
-g_{\mu\nu} \left[ \frac{1}{2} g^{\alpha\beta} 
\left( \Phi_{, \, \alpha}^*\Phi_{, \, \beta}+
\Phi_{, \, \beta}^*\Phi_{, \, \alpha} \right)+U(\left|\Phi\right|)\right]
 \ .
\end{eqnarray}
Variation with respect to the scalar field
leads to the matter field equation,
\begin{eqnarray}
\label{scalar-eq}
\frac{1}{\sqrt{-g}} \partial_\mu \big(\sqrt{-g} \partial^\mu\Phi \big)=\frac{\partial U}{\partial\left|\Phi\right|^2} \Phi.
\end{eqnarray} 
The potential $U$ can be decomposed according to
\begin{eqnarray}
\label{pot1}
U(|\Phi|)= \mu^2 |\Phi|^2 +U_{int}(|\Phi|),
\end{eqnarray}
 where $\mu$ is the mass of the field, and $U_{int}$ is a self-interaction potential.
As discussed in \cite{Astefanesei:2003qy} for $\mu^2>0$, this model possesses finite mass solutions  
even in the absence of self-interaction,
$U_{int}=0$, the so-called 'mini-boson stars'.

However, the inclusion
of an interaction potential may lead
to a more complex picture (see $e.g.$ \cite{Colpi:1986ye} for the $\Lambda=0$ case). 
Our choice of  $U_{int}$ was guided by the requirement that nontopological solitons
exist also in a fixed AdS background.
Moreover, we are interested in the case when the solutions would possess a nontrivial static limit. 
As discussed in \cite{Hertog:2004rz}, this 
requires the occurrence of negative energy densities,
 $i.e.$ $U(|\Phi|)<0$
in some region.
Since, in order 
to make contact with the previous work on mini-boson stars, we restrict the numerical part of our
study to the case $\mu^2>0$, this implies that  $U_{int}$ is not strictly positive. 

We have found that the simplest choice of the interaction potential 
satisfying these conditions is $U_{int}=-\lambda|\Phi|^{2k}$, with $k>1$ and $\lambda>0$.
Most of the  results in this work are found\footnote{Note that the  exact solution with spherical symmetry
 in Section 3 covers a more general range of $k$.} for $k=2$,
such that 
\begin{eqnarray}
\label{U}
 U(|\Phi|)=\mu^2 |\Phi|^2  -\lambda |\Phi|^{4}.
\end{eqnarray} 
Although the action (\ref{action}) together with (\ref{U}) does not seem to correspond to any supergravity 
model, it is likely that some features of its solutions are generic.
 In particular, we have found the same general picture 
 for a more general potential (which was used in the previous studies 
 \cite{Volkov:2002aj},
\cite{Kleihaus:2005me}, 
\cite{Kleihaus:2007vk}
 on $\Lambda=0$ spinning Q-balls and boson stars)
 \begin{eqnarray}
\label{Un}
 U(|\Phi|)=\mu^2 |\Phi|^2 -\lambda |\Phi|^{4}+\nu |\Phi|^{6},
\end{eqnarray} 
 provided that
 the new coupling constant $\nu>0$ is  small enough.

\subsection{The Ansatz }

We are interested in stationary axially symmetric configurations, 
with a spacetime geometry admiting two Killing vectors 
$\partial_t$
and 
$\partial_{\varphi}$,
in a system of adapted coordinates $\{t, r, \theta, \varphi\}$.
Thus the line element can be written as
\begin{eqnarray}
\label{ansatzg}
ds^2 =- F_0 N dt^2 
+ F_1 \left( \frac{dr^2}{N} + r^2 \, d\t^2 \right) 
  + F_2 r^2 \sin^2 \t  \left( d \varphi
- \frac{W}{r}   dt \right)^2   . 
\end{eqnarray}
The  metric functions $F_0$, $F_1$, $F_2$ and $W$
depend on the variables $r$ and $\theta$ only, while
\begin{eqnarray}
N=1+\frac{r^2}{\ell^2}
\end{eqnarray}
is a suitable 'background' function.
 

For the scalar field $\Phi$ we adopt the stationary ansatz 
\begin{eqnarray}
\label{ansatzp}
\Phi (t,r,\t, \varphi)= \phi (r, \t)
 e^{ i( n \varphi-\omega  t )}   , 
\end{eqnarray}
where $\phi (r, \theta)$ is a real function,
and $\omega $ and $n$ are real constants.
Single-valuedness of the scalar field requires
$\Phi(\varphi)=\Phi(2\pi + \varphi)$;
thus the constant $n$ must be an integer,
$i.e.$, $n \, = \, 0, \, \pm 1, \, \pm 2, \, \dots$~.
In what follows, we shall take $n\geq 0$ and $\omega \geq 0$,
without any loss of generality.

 The spherically symmetric limit is found for $n=0$, in which case the
 functions $F_0,F_1,F_2$ and $\phi$
 depend  only on $r$, with $F_1=F_2$ and $W=0$. 
 
\subsection{The boundary conditions}
The solutions in this work describe horizonless, particle-like configurations. 
A study of an approximate form of the solutions as a power series around $r=0$ leads to the following
 boundary conditions at the origin\footnote{For spherically
symmetric solutions, the scalar field is nonvanishing ar $r=0$.}:
\begin{eqnarray}
\label{bc0} 
\partial_r F_i|_{r=0}=0, ~~
W|_{r=0}=0,~~
\phi| _{r =0}=0~,
\end{eqnarray}
 (with $i=0,1,2$).
At infinity, the AdS background is approached, while the scalar field vanishes.
Without any loss of generality,
we are choosing a frame in which the solutions do not rotate at infinity,
the conformal boundary being a static Einstein universe $R\times S^2$.
The boundary conditions compatible with these requirement are
\begin{eqnarray}
\label{bcinf} 
F_i|_{r \rightarrow \infty} =1,~~
W|_{r \rightarrow \infty} =0, ~~
\phi| _{r \rightarrow \infty}=0 \ .
\end{eqnarray}
For $\t=0,\pi$  
we require the boundary conditions
\begin{eqnarray}
\label{bct0} 
\partial_{\t} F_i|_{\t=0,\pi}=0, ~~
\partial_{\t} W |_{\t=0,\pi}=0,~~
\f |_{\t=0,\pi}=0.
\end{eqnarray}
 The absence of conical singularities
  imposes on the symmetry axis the supplementary condition 
$F_1|_{\theta=0,\pi}=F_2|_{\theta=0,\pi},$
which is  used to verify the accuracy of the solutions.

Also, all solutions in this work 
are invariant under the parity transformation $\theta \to\pi-\theta $.
We make use of this symmetry to integrate the equations for $0\leq \theta\leq \pi/2$ only, the
following boundary conditions being imposed in the equatorial plane
\begin{eqnarray}
\label{bctpi2} 
\partial_{\t} F_i|_{\t=\pi/2}=0 ,~~
\partial_{\t} W |_{\t=\pi/2}=0 \ ,~~
\partial_{\t} \phi |_{\t=\pi/2}=0 \ .
 \end{eqnarray}

\subsection{The far field asymptotics and global charges}
For solutions
with $\mu^2>0$ (the only case considered in the numerics),
the scalar field decays asymptotically as 
\begin{eqnarray}
\label{asym-scalar0}
\phi\sim 
\frac{c_1(\theta)}{r^{\frac{3}{2}\left( 1+\sqrt{1+\frac{4}{9}\mu^2 \ell^2} \right)}}+\dots~.
\end{eqnarray}
The physical interpretation of $c_1(\theta)$
is that it corresponds, up to a normalization,
to the expectation value of some scalar operator in the dual theory.

Without entering into details, 
we mention that the picture is more complicated \cite{Henneaux:2006hk} if one allows for a tachyonic
mass of the scalar field, $\mu^2<0$.
For $-9/4< \mu^2\ell^2<-5/4$,
the general asymptotic behaviour of the scalar field 
is more complicated\footnote{For
a field which saturates the Breitenlohner-Freedman bound $\mu^2\ell^2=-9/4$,
one finds $\Delta_+=\Delta_-=\Delta$, and the second solution asymptotically behaves 
like $\log r/r^\Delta$.
},  with the existence of a second mode apart from (\ref{asym-scalar0}): 
\begin{eqnarray}
\label{asym-scalar}
\phi\sim 
\frac{c_1(\theta)}{r^{\Delta_+}}+
\frac{c_2(\theta)}{r^{\Delta_-}},
\end{eqnarray}
where
\begin{eqnarray}
\label{Deltapm}
\Delta_{\pm}=\frac{3}{2}\left( 1\pm \sqrt{1+\frac{4}{9}\mu^2 \ell^2} \right). 
\end{eqnarray}
For this range of $\mu^2<0$, both modes above are normalizable in the sense that the spatial integral of $j^t$ is finite, $i.e.$
the scalar field possesses a finite Noether charge.
To have a well defined theory,
one must specify a boundary condition at infinity, $i.e.$
to choose a relation between $c_1$ and $c_2$, the standard choice being $c_2=0$. 
However, as discussed $e.g.$ in  \cite{Hertog:2004dr}, \cite{Henneaux:2004zi},
the solutions with a slower decay at infinity, $c_2 \neq 0$,
are also physically acceptable, the AdS charges involving in this
case a scalar field contribution (thus the 
expression (\ref{EJ}) below would not be valid).

Restricting to the case $\mu^2>0$, the scalar field decays asymptotically faster than $1/r^3$ and
thus the Einstein equations (\ref{Einstein-eqs}) imply the following form of the metric functions
as $r\to \infty$
 \begin{eqnarray}
\nonumber
&&F_0=1+ \frac{f_{03}(\theta)}{r^3}+O(1/r^5),
~~~F_1=1+ \frac{f_{13}(\theta)}{r^3}+O(1/r^5), 
\\
\label{asym1}
&&F_2=1+ \frac{f_{23}(\theta)}{r^3}+O(1/r^5),
~~W=\frac{w_2(\theta)}{r^2}+O(1/r^4),
\end{eqnarray}
in terms of two functions $f_{13}(\theta)$ and $w_2(\theta)$ which result from the numerics, 
with 
\begin{eqnarray}
f_{03}(\theta)=-3 f_{13}(\theta)-\frac{4}{3}\tan\theta f_{13}'(\theta),~~{\rm and}~~
f_{23}(\theta)=  f_{13}(\theta)+\frac{4}{3}\tan\theta f_{13}'(\theta).
\end{eqnarray} 

A straightforward computation based on the formalism in \cite{Balasubramanian:1999re} 
leads to the following expression for 
the mass-energy $E$ and angular momentum $J$ of the  configurations
\begin{eqnarray}
\label{EJ}
E= \frac{1}{8 G \ell^2}\int_{0}^\pi d\theta \sin\theta \bigg(5 f_{13}(\theta)+3 f_{23}(\theta)\bigg),~~
J=-\frac{3}{8 G }\int_{0}^\pi d\theta \sin^3\theta ~w_2(\theta) .
\end{eqnarray} 
(Note that the same result can be derived by using
 the Ashtekar-Magnon-Das conformal mass definition \cite{Ash}).
 Moreover, 
the same expression for the angular momentum is found from the Komar integral:
\begin{eqnarray}
\label{j1}
 &J=\frac{1}{8 \pi G }\int R_{\varphi}^t \sqrt{-g} dr  d\theta d\varphi
 =
%
  -\frac{1}{8 G }\int_{0}^{\infty}dr \int_0^\pi d \theta
 \bigg [
 \bigg(r^4 \sqrt{\frac{F_2^4}{F_0}}\sin^3 \theta   \big(\frac{W}{r} \big)_{,r} \bigg)_{,r}
 + \bigg(\frac{r}{N} \sqrt{\frac{F_2^3}{F_0}}\sin^3 \theta   W_{,\theta} \bigg)_{,\theta}
 \bigg ].~{~~}
\end{eqnarray} 
The solutions possess also a conserved Noether charge ($i.e.$ the total
particle number)
\begin{eqnarray}
\label{Q1}
Q= \int j^t \sqrt{-g} dr  d\theta d\varphi=
2 \pi \int_0^\infty dr \int_0^\pi d\theta~
2r \sin \theta F_1\sqrt{\frac{F_2}{F_0}}
\frac{ \phi^2}{N}(\omega r-n W).
\end{eqnarray} 
Since $T_\varphi^t=n j^t$, from (\ref{j1}) 
and the Einstein equation $R_{\varphi}^t=8\pi G T_{\varphi}^t$
we find that the generic relation
\begin{eqnarray}
\label{JQ}
J= n Q~,
\end{eqnarray} 
(which was proven in \cite{Schunck:1996wa}
 for asymptotically flat spinning solutions)
holds also in the AdS case.
As a result, the spinning solitons
do not emerge as perturbations
of the spherically symmetric
configurations, $i.e.$
there are no slowly rotating
solutions in this model (note also that the expression of the scalar field
potential was not used in the derivation of (\ref{JQ})). 

These scalar solitons have no horizon and therefore they are zero entropy objects,
without an intrinsic temperature.
The first law of thermodynamics
reads   in this case  \cite{Lee:1991ax}
$dE= \omega dQ=\frac{\omega }{n} dJ$.

\subsection{The numerical scheme}
There are not  many studies on numerical 
solutions of Einstein gravity with negative cosmological constant
coupled with  matter fields,
 describing stationary, axially symmetric configurations.
The model in this work provides perhaps 
the simplest test ground for 
investigating various approaches and developing numerical techniques for elliptic 
problems with AdS asymptotics\footnote{Because the ansatz (\ref{ansatzp}) has an explicit dependence 
 on both $\varphi$ and $t$,
 the scalar field is neither static nor axisymmetric.
 However, all physical quantities, such as the current $j^\mu$
 and the energy-momentum tensor $T_{\mu\nu}$,
 will exhibit no dependence on  $\varphi$ and $t$.}.

The solutions of the field equations (\ref{Einstein-eqs}), (\ref{scalar-eq}) 
are found  by using an approach originally proposed in \cite{Kleihaus:1996vi}
for $d=4$ asymptotically flat solutions of Einstein 
gravity coupled with Yang-Mills gauge fields. 
The equations for the metric functions  $ \mathcal F=(F_0,F_1,F_2,W)$ 
 employed in the numerics,
are found by using a  suitable combination of the Einstein equations (\ref{Einstein-eqs}),
$E_t^t =0,~E_r^r+E_\theta^\theta =0$,  $E_\varphi^\varphi=0$
and  $E_{\varphi}^{t} =0$,
 which diagonalizes them $w.r.t.$ $\nabla^2 \mathcal F$ 
 (where $\nabla^2=\partial_{rr}+\frac{1}{r}\partial_{r}+\frac{1}{r^2 N}\partial_{\theta\theta}$).
 An important issue here concerns the status of the 
remaining equations $E_\theta^r =0,~E_r^r-E_\theta^\theta  =0$,
 which
yield two constraints. 
Following \cite{Wiseman:2002zc}, one can show that
the identities $\nabla_\mu E^{\mu r} =0$ and $\nabla_\mu E^{\mu \theta}=0$, 
imply the Cauchy-Riemann relations
$
\partial_{\bar r} {\cal P}_2  +
\partial_\theta {\cal P}_1  
= 0 ,~~
 \partial_{\bar r} {\cal P}_1  
-\partial_{\theta} {\cal P}_2
~= 0 ,
$
with ${\cal P}_1=\sqrt{-g} E^r_\theta$, ${\cal P}_2=\sqrt{-g}r \sqrt{N}(E^r_r-E^\theta_\theta)/2$
and $d\bar r=\frac{dr}{r \sqrt{N}}$.
Therefore the weighted constraints still satisfy Laplace equations, and the constraints 
are fulfilled, when one of them is satisfied on the boundary and the other 
at a single point
\cite{Wiseman:2002zc}. 
From the boundary  conditions (\ref{bc0})-(\ref{bct0}) we are imposing,
it turns out that this is the case for our solutions,
 $i.e.$ the numerical scheme is consistent.
 
To obtain spinning boson star solutions,
we solve  numerically the set of five coupled non-linear
elliptic partial differential equations for $(\mathcal F,\phi)$,
subject to the  boundary conditions (\ref{bc0})-(\ref{bct0}).
We employ a compactified radial coordinate $\bar r=  r/(1+ r)$
 which maps spatial infinity to the finite value $\bar r=1$.
 Then the equations are discretized on a non-equidistant grid in
$\bar r$ and $\theta$.
Typical grids used have sizes $250 \times 30$,
covering the integration region
$0\leq \bar r \leq 1$ and $0\leq \bar \theta \leq \pi/2$.
(See \cite{Kleihaus:1996vi} and \cite{schoen}  
for further details and examples for the numerical procedure.) 
The numerical calculations are based on the Newton-Raphson method
and are performed with help of the  software package FIDISOL \cite{schoen},
which provides also an error estimate for each unknown function.
The typical relative error for the solutions 
in this work is smaller that
$10^{-3}$.
 
\section{The solutions in the probe limit}

We shall start with a discussion of the solutions in a fixed AdS
background, $i.e.$ 
without backreaction, $F_0=F_1=F_2=1,~W=0$ in the metric ansatz (\ref{ansatzg}).
The problem is much easier to study in this limit and
the solutions exhibit already some basic features of
the gravitating configurations.

The usual Derick-type scaling argument (see $e.g.$
the discussion in \cite{Radu:2008pp} 
for the $\Lambda=0$ limit) implies that
the Q-balls in a fixed AdS background satisfy
the following virial identity
\begin{eqnarray}
\label{virial}
\int_0^\infty dr \int_0^\pi  d\theta \sin \theta
\bigg [
N\phi_{,r}^2
+\frac{\phi_{,\theta}^2}{r^2}
+\frac{n^2\phi^2}{r^2\sin^2\theta}
+3\left(U(\phi)-\frac{\omega^2\phi^2}{N}\right)
+\frac{2r}{\ell^2}\left(\phi_{,r}^2+\frac{ \omega^2 \phi^2}{N^2}\right)
\bigg]=0,
\end{eqnarray}
which was used as a further test of the numerical accuracy.
It is clear that the 
solutions with a strictly positive potential, $U(\phi)>0$,
owe their existence to the harmonic time dependence of the scalar field.
Also, the solutions may exist in the $ \omega\to 0$ limit 
as long as $U(\phi)$ is allowed to take negative values.

As usual in the absence of back reaction, 
the total mass-energy $E$ and angular momentum of the configurations
are found by integrating over the entire space
the components $-T_t^t$ and $T_\varphi^t$ of the energy 
momentum tensor.

%

\subsection{Spherically symmetric configurations.
An exact solution}

In the spherically symmetric limit, the equation (\ref{scalar-eq}) with a self-interaction
potential $U_{int}=-\lambda \phi^{2k}$
admits the following simple exact solution, which to our knowledge,
was not yet discussed in the literature: 
\begin{eqnarray}
\label{ex-sol}
\Phi(r,t)=\left(
\frac{\mu^2}{\lambda}\frac{\Delta^2-\ell^2  \omega^2}{(\Delta-3)(\Delta+1))}
\right)^{\frac{\Delta}{2}} {\left(1+\frac{r^2}{\ell^2} \right)^{-\frac{\Delta}{2}}} e^{-i  \omega t}, 
\end{eqnarray}
with $\Delta=\Delta_{\pm}$, as  given by (\ref{Deltapm}).

For this exact solution, 
the coefficient $k$ of the self-interaction term 
in the scalar field potential (\ref{U}) is fixed by $\Delta$,
\begin{eqnarray}
\label{k}
k=1+\frac{1}{\Delta}.
\end{eqnarray}
This is a one parameter family of solutions which, for given input parameters $\mu^2,\lambda$ and $\ell$,
can be parametrized by the frequency of the field.
Choosing $\Delta=\Delta_{+}$ leads to a range $1<k<5/3$ for the parameter $k$ in the self-interaction potential, 
the solutions possessing in this case a finite mass-energy
and Noether charge:
\begin{eqnarray}
&&
E= V_2 \sqrt{\pi} \ell \left(\frac{\mu^2}{\lambda} \right)^\Delta 
\frac{\Gamma(\Delta-\frac{1}{2})}{\Gamma(\Delta+2)}
\left( \frac{\Delta^2-\ell^2 \omega^2}{(\Delta-3)(\Delta+1)} \right)^\Delta
\left( \Delta^2+ \omega^2\ell^2(2\Delta+1) \right),
\\
\nonumber
&&
Q=V_2\sqrt{\pi}\frac{\ell^3  \omega}{2}\left(\frac{\mu^2}{\lambda} \right)^\Delta 
\frac{\Gamma(\Delta-\frac{1}{2})}{\Gamma(\Delta+1)}
\left( \frac{\Delta^2-\ell^2  \omega^2}{(\Delta-3)(\Delta+1)} \right)^\Delta~, 
\end{eqnarray}
 with $V_2=4\pi$ the area of two-sphere. 

\setlength{\unitlength}{1cm}
\begin{picture}(8,6) 
\put(-0.5,0.0){\epsfig{file=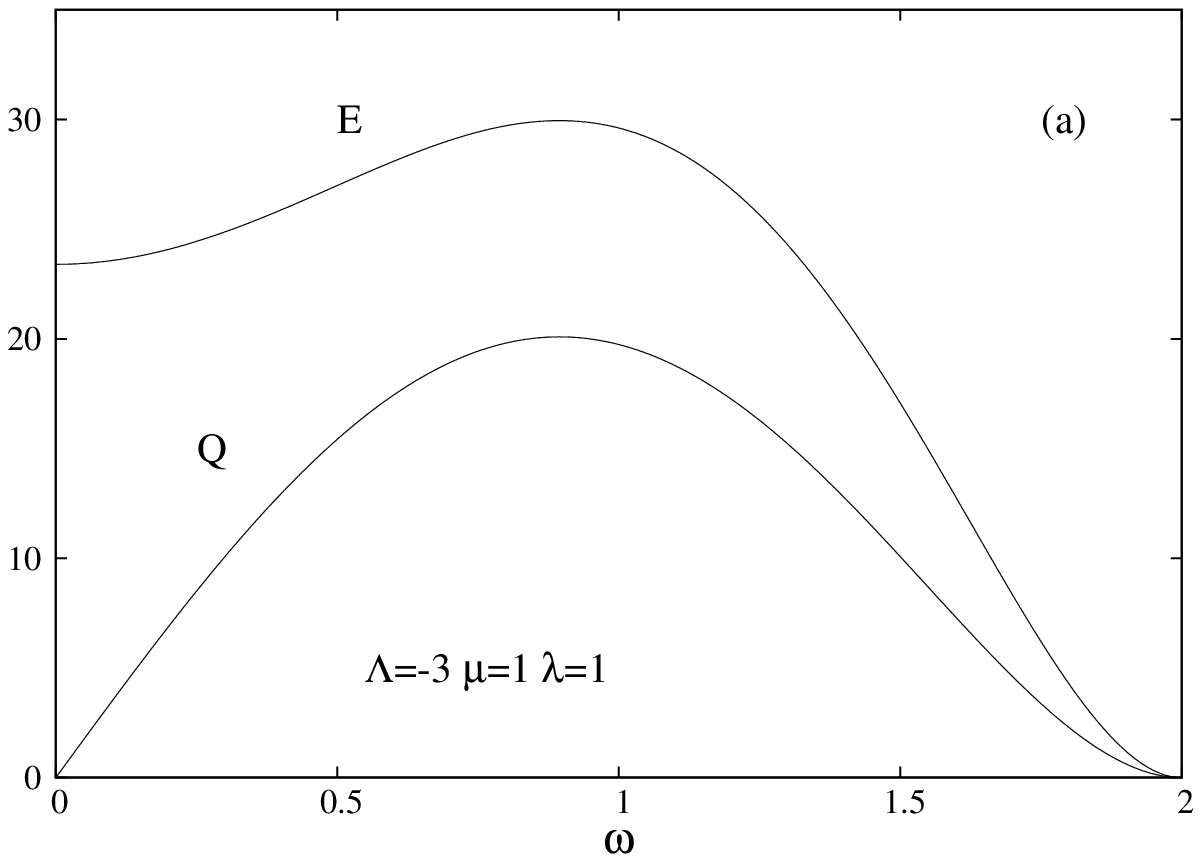,width=8cm}}
\put(8,0.0){\epsfig{file=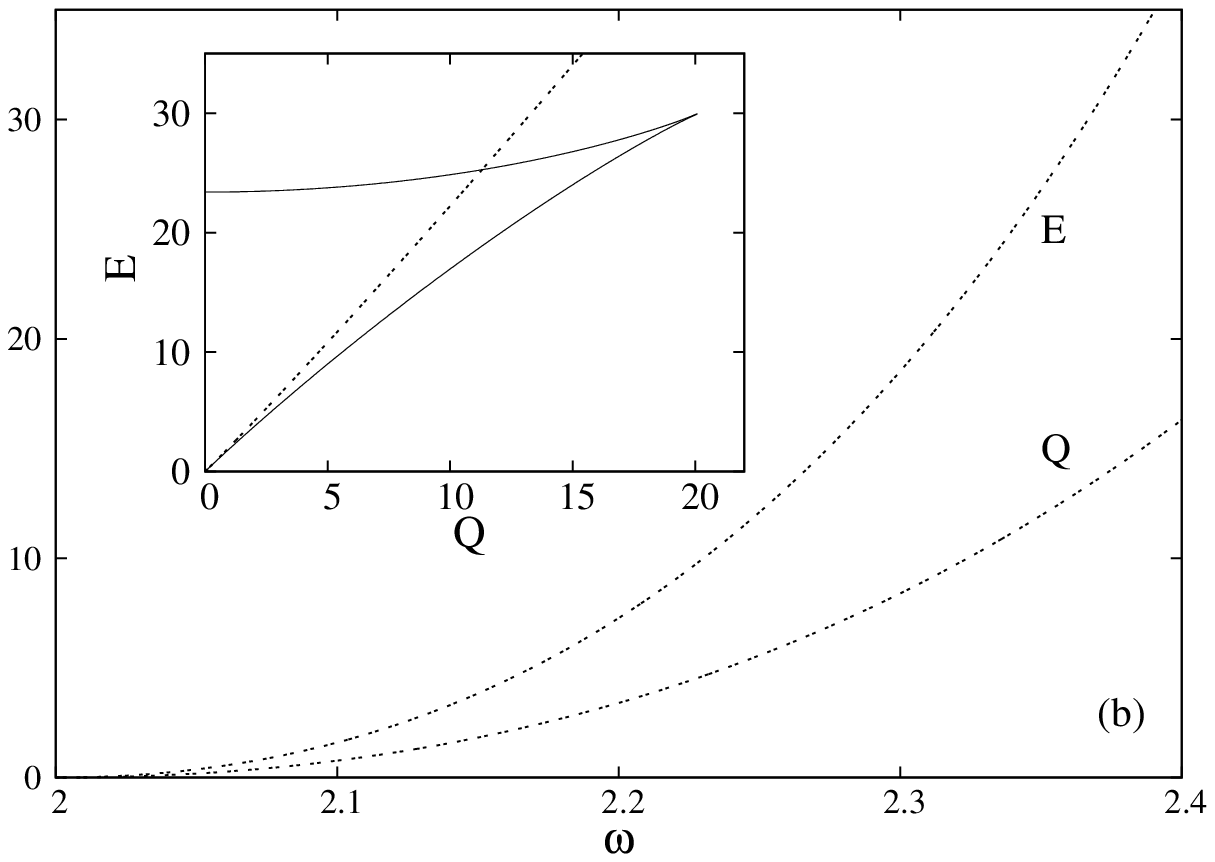,width=8cm}}
\end{picture}
\\
\\
{\small {\bf Figure 1.} The mass-energy and Noether charge of the 
exact solution (\ref{ex-sol}) with a $\phi^3$-self interaction potential are shown as a function of
frequency.
   }
\vspace{0.5cm}
 %
 \\
 
  An interesting case here corresponds to $k=3/2$, $i.e.$ 
an $(\Phi^*\Phi)^{3/2}$ interaction term\footnote{A $U(1)$-symmetric
cubic interaction $(\Phi^*\Phi)^{3/2}$ in the potential
is not unusual in the Q-ball literature, see $e.g$
\cite{Kusenko:1997ad}, \cite{Tamaki:2010zz}, such terms 
occurring in the minimal supersymmetric standard model.}
in the  potential (\ref{U}).
This case is realized for a scalar field mass $\mu^2=-{2}/{\ell^2}$  and a choice $\Delta=\Delta_+=2$.
The total mass-energy and Noether charge of this solution
are given by 
\begin{eqnarray}
E= V_2 
\frac{\pi}{108\lambda^2\ell^3}(4-\ell^2  \omega^2)^2
(4+5\ell^2  \omega^2),~~
Q= V_2 \frac{\pi  \omega (4-\ell^2  \omega^2)^2}{18 \lambda^2 \ell},
\end{eqnarray}
being shown in Figure 1 as a function of the frequency.
 
 Remarkably, the  picture in Figure 1a 
 has been recovered for most of the numerical solutions with a self-interaction potential and $\mu^2>0$
in this work (including the spinning ones with $n\neq 0$).
 The 
occurrence of an extra-branch of solutions with $ \omega> \omega_c$ (see Figure 1b)
is a feature of the $\phi^3$ potential.
(Note that this branch does not appear for the exact solution with a $\phi^4$ potential.)

For the same value of the scalar field mass, when choosing instead
$\Delta=\Delta_-=1$ in (\ref{ex-sol}), (\ref{k}), one recovers the model with
a $\phi^4$ potential.   
Unfortunately, the total mass-energy of these solutions, as defined
in the usual way as the integral of $T_t^t$,
diverges linearly\footnote{However, the mass can be regularized
by supplementing the action with a suitable scalar field boundary counterterm.},
$E=E_{div}+E_0$, with $E_{div}=- V_2\frac{ (1-\ell^2  \omega^2) }{2\lambda \ell}r_c$ 
(where $r_c\to \infty$), 
while the charge is finite:
\begin{eqnarray}
\label{phi4}
E_0=V_2\frac{\pi (1-\ell^2  \omega^2)(1+3 \ell^2  \omega^2)}{16\lambda \ell},~~
Q=V_2\frac{\pi \ell  \omega(1- \omega^2 \ell^2)}{4 \lambda}.
\end{eqnarray}

Solutions beyond the framework (\ref{ex-sol}), (\ref{k})
are found by using a numerical approach,
for a generic ansatz $\Phi = \phi (r) e^{-i  \omega t} $. 
Here,  for  simplicity we restrict ourselves to the case of nodeless solutions.
Then, for $\mu^2>0$ and a self-interaction
potential $U_{int}=-\lambda \phi^{4}$,
it turns out that the picture in Figure 1a is generic.
For any $\Lambda$, the solutions exist for  a limited range of frequencies,
$0\leq  \omega< \omega_c= \Delta_+/\ell$.
The mass-energy and Noether charge are bounded and approach a maximum
for some $ \omega$ around $ \omega_c/2$.
The same pattern is recovered  
 in the presence of an extra $\nu \phi^6$ self-interaction term (with $\nu>0$), provided that
 the new coupling constant is small enough.

\subsection{Spinning scalar solitons in an AdS background}

The spinning generalizations of these solutions are found by taking $n\neq 0$
in the general ansatz (\ref{ansatzp}).
For a $\phi^4$ self-interaction potential,
apart from the winding number $n$,
the input parameters are $\ell$, $ \omega$, $\mu$ and $\lambda$.
\\
\setlength{\unitlength}{1cm}
\begin{picture}(8,6) 
\put(-0.5,0.0){\epsfig{file=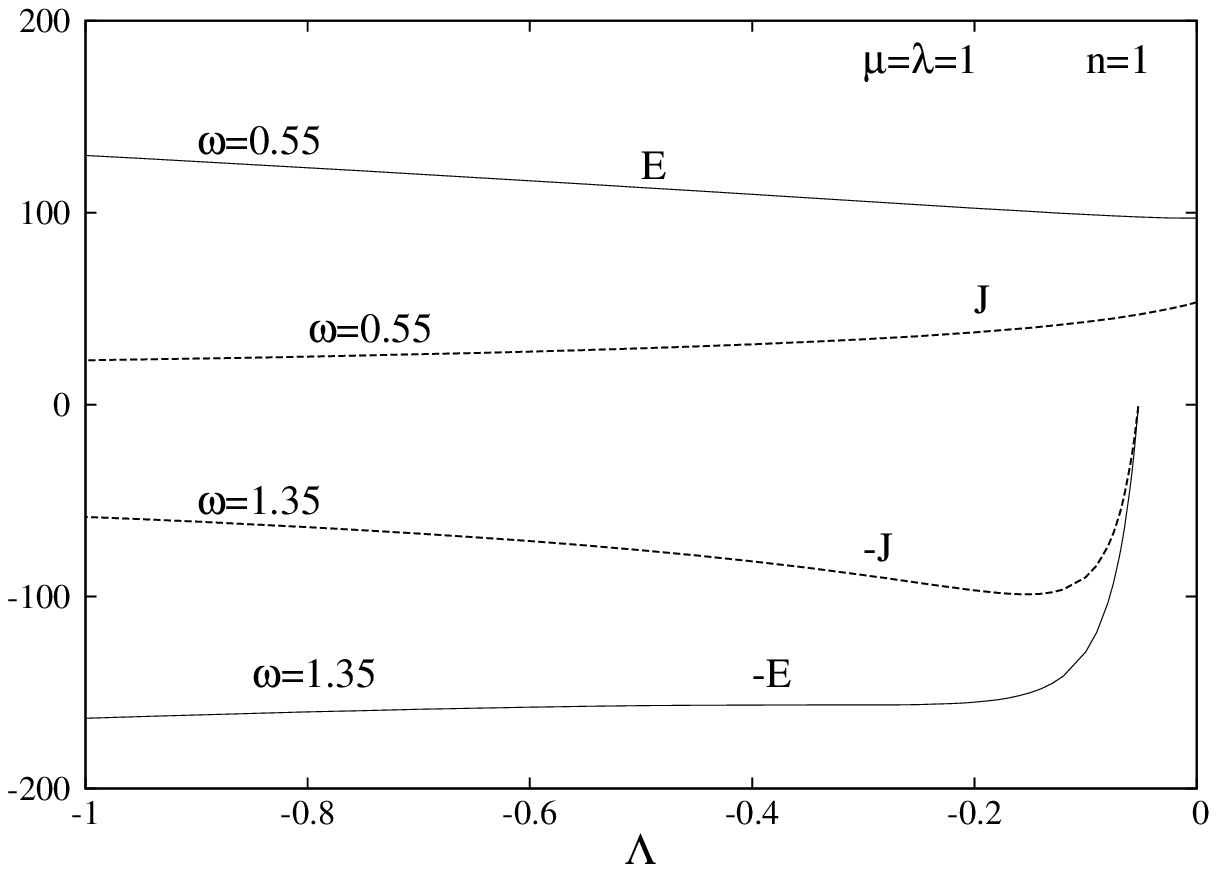,width=8cm}}
\put(8,0.0){\epsfig{file=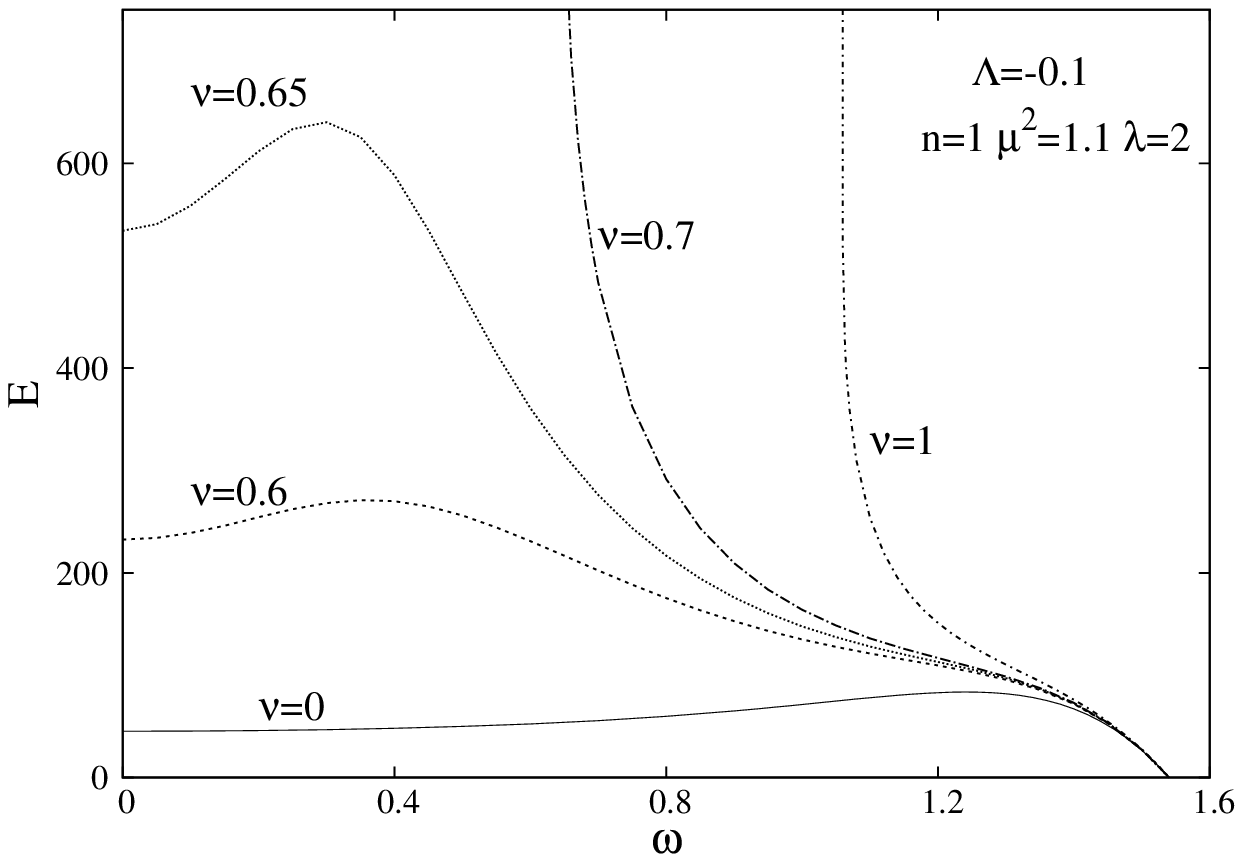,width=8cm}}
\end{picture}
\\
\\
{\small {\bf Figure 2.} {\it Left:} The mass-energy $E$ and angular momentum $J$
 of $n=1$ spinning  nongravitating Q-ball solutions 
 are shown as a function of the cosmological constant $\Lambda$ for two values of
 the frequency $ \omega$.
  {\it Right:}
  The mass-energy $E$  
 of $n=1$ spinning  nongravitating Q-ball solutions with a potential 
 $U(\phi)=\mu^2 \phi^2-\lambda \phi^4+\nu \phi^6$
 are shown as a function of the frequency for several values of the coefficient of the
 $\phi^6$ term.
   }
\vspace{0.5cm}
 \\
However, the model admits the scaling symmetry
$r \to r/\mu$, $ \omega\to  \omega \mu$, $\ell \to \ell/\mu$,
together with the scalar field redefinition $\phi\to \phi/c$, $\lambda\to \lambda c^2$,
which allows us to set $\mu=\lambda=1$ without any loss of generality.

The following picture appears to be generic for the numerical solutions in this work:
first, no solutions exist 
for frequencies above a critical value $ \omega= \omega_c$.
As $\omega\to \omega_c$, the solution emerges 
 as a perturbation around the ground state $\phi=0$, with
\begin{eqnarray}
\label{pert2}
\delta \phi(r,\theta) \simeq \frac{(r \sin\theta)^n}{\left(1+\frac{r^2}{\ell^2}\right)^{\frac{1}{2} \omega_c \ell}} 
\end{eqnarray}
being the regular solution of the linearized Klein-Gordon equation in a fixed AdS background
(here we restrict our discussion to nodeless configurations).
The critical frequency is given by
\begin{eqnarray}
\label{pert1}
  \omega_{c}=\frac{n+\Delta_{+}}{\ell},
\end{eqnarray}
being found by requiring the perturbation $\delta \phi(r,\theta)$
to be regular at both $r\to 0$ and $r\to \infty$. 
This upper bound on the frequency
is universal, 
and holds also in the presence of gravity.

When decreasing the frequency, the nonlinear term in $\phi$
starts to contribute
 and the mass-energy and the angular momentum of the 
solitons start to increase.
Both $E$ and $J$ approach a maximum at some intermediate value of $ \omega$,
decreasing afterwards.
As $ \omega\to 0$, 
the solutions describe finite mass, static, axially symmetric (for $n\neq 0$) solitons, 
though
with a vanishing Noether charge.

The dependence of the solutions on the cosmological constant is shown in Figure 2 (left).
One can see that the pattern depends on the value of the frequency.
For $\omega>\mu$, the solutions stop to exist for a maximal value of the cosmological constant,
 $\Lambda=-\frac{4(\mu^2-\omega^2)^2}{ 3( \omega(1+2n/3)+\sqrt{ \omega^2+4\mu^2n(n+3)/9})^2}$ 
 (which results from (\ref{pert1})),
 where both $E$ and $Q$ vanish  in that limit.
For $\omega<\mu$,
 one finds solutions for all values of cosmological constant, including $\Lambda=0$.

  We have also constructed AdS generalizations of the flat spacetime Q-balls
 with the usual potential choice (\ref{Un}).
As one can see in Figure 2 (right), the picture found
for the $\phi^4$-model is recovered for small enough values of  $\nu$.
However, for $\nu$ above a critical value (around $0.67$ for those parameters),  the solutions exist
only for $  \omega_{min}< \omega< \omega_c$ (with $\omega_{min}>0$).
Similar to the flat space solutions \cite{Kleihaus:2005me}, both the 
mass-energy and Noether charge diverge
as $ \omega\to  \omega_{min}$.

\setlength{\unitlength}{1cm}
\begin{picture}(8,6) 
\put(-0.5,0.0){\epsfig{file=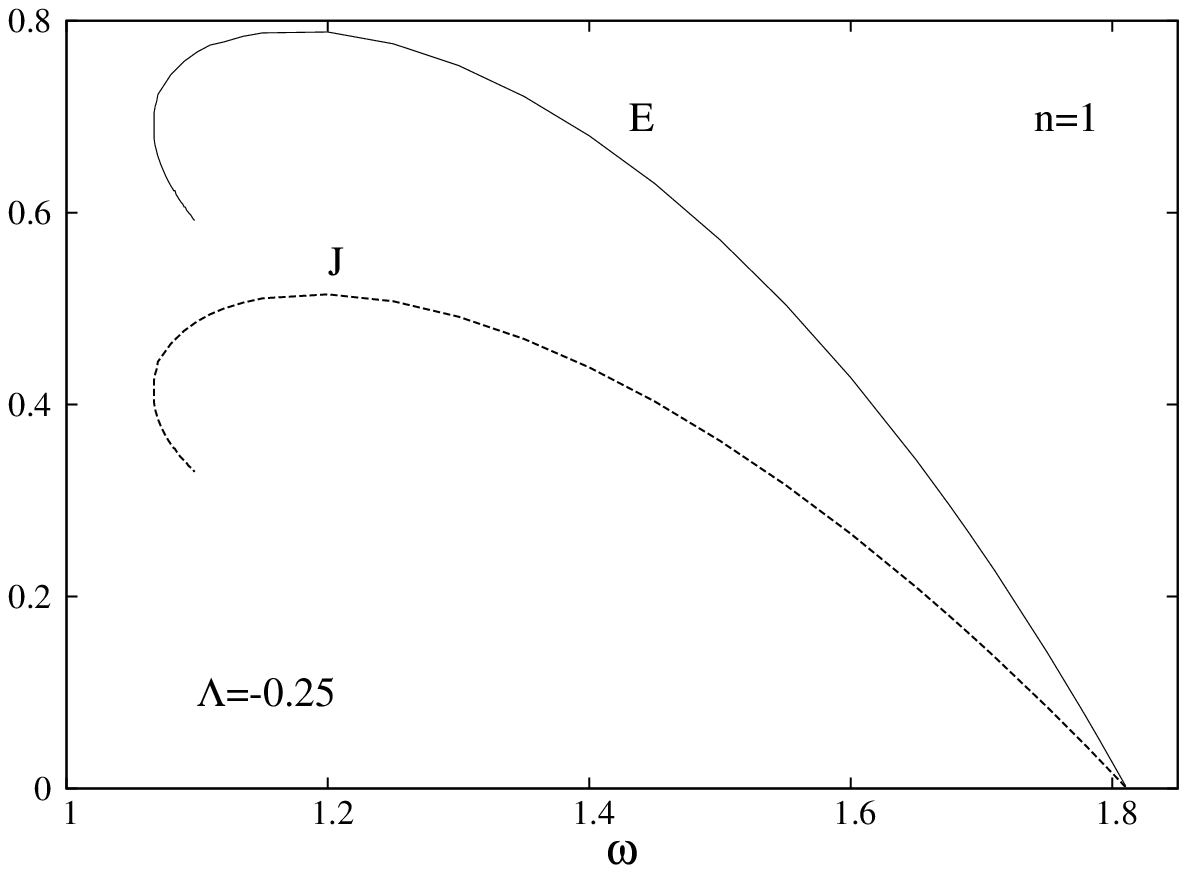,width=8cm}}
\put(8,0.0){\epsfig{file=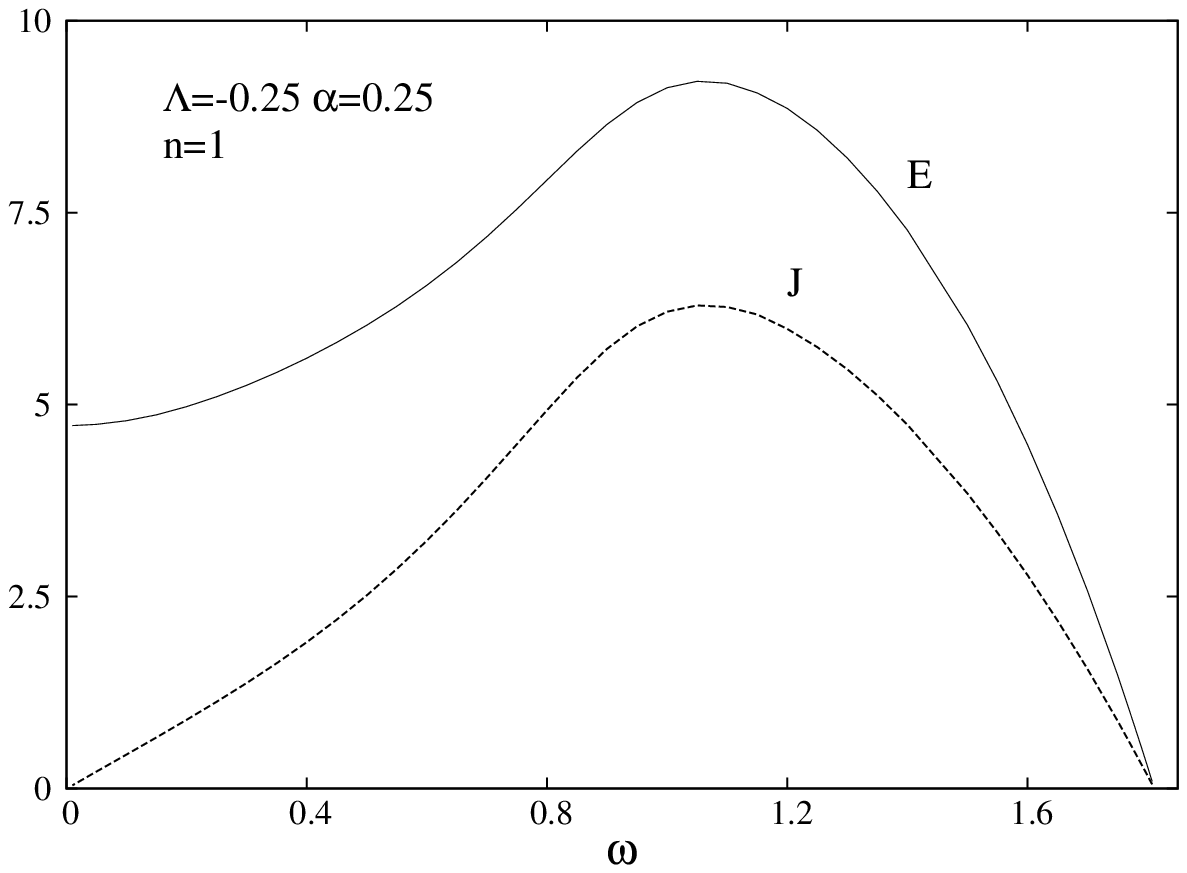,width=8cm}}
\end{picture}
\\
\\
{\small {\bf Figure 3.} 
The mass-energy $E$ and the angular momentum $J$ are shown as a function of the frequency
$\omega$
for (mini-)boson star solutions without a self-interaction potential (left)
and for gravitating Q-balls with a $\phi^4$ interaction potential (right).
   }
\vspace{0.1cm}

\section{Gravitating spinning scalar soliton}

We have found that all Q-balls in a fixed AdS background 
allow for gravitating generalizations.
Moreover, nontrivial solutions are found in this case
even in the absence of a self-interaction term in the scalar potential (\ref{U}),
generalizing for $J\neq 0$ the static AdS
boson stars in \cite{Astefanesei:2003qy}.

Let us start with a brief discussion of those configurations with  
  $U=\mu^2 \phi^2$ 
(usually called
mini-boson stars in the literature), 
which play an important role in the limiting behaviour of 
the configurations with a $\phi^4$ term in the potential.
Restricting again to a real mass of the scalar field, $\mu^2>0$, the usual 
rescaling $r\to r\mu$ implies that the model depends on two
dimensionless parameters, $\ell \mu$ and $ \omega/\mu$, only, the factor $1/\sqrt{8\pi G}$
being absorbed in $\phi$.  
Our results show that similar to the spherically symmetric case,
the spinning solutions exist for a limited range of frequencies
$0< \omega_{min}< \omega< \omega_c$,  emerging
as a perturbation of AdS spacetime 
for a critical frequency $ \omega_c$ as given by (\ref{pert1}). 
Thus, as expected, the value $ \omega=0$ is not approached in the absence of a 
scalar field self-interaction\footnote{Finding
AdS
static solitons requires a violation of the energy conditions \cite{Hertog:2004rz}, which is not the case for
the boson stars  with $U(|\Phi|)=\mu^2 |\Phi|^2>0$.}.
As $ \omega\to  \omega_{min}$,
a backbending towards larger values of $ \omega$ is observed, see Figure 3 (left).
We conjecture that, similar to the spherically symmetric case, 
this backbending  would lead to an inspiraling of the solutions
towards a limiting configuration with $ \omega_0> \omega_{min}$.
Note also that the mass-energy and the angular momentum of these mini-boson star solutions stay finite
in the allowed range of frequencies.

The picture is more complicated for solitons
with a self-interaction term in the potential.
For simplicity, we shall restrict\footnote{However, we have also constructed  
 gravitating solutions with a $\phi^6$-term in the potential,
in which case we did not find new qualitative features.} our discussion to the case of 
a $\phi^4$ potential (\ref{U}).
Without any loss of generality,
 one can set  $\mu=\lambda=1$ for the two parameters in (\ref{U}).
This choice is achieved by using the rescaling 
$r \to  r/\mu$, $\ell \to  \ell/\mu  $, $\omega \to \omega \mu$, 
together with a redefinition of the scalar field
$\phi \to \phi \mu/\sqrt{\lambda}$.
This  reveals the existence of a dimensionless parameter $\alpha^2=4 \pi G \mu^2/\lambda$,
such that the
Einstein equations read $R_{ij}-\frac{1}{2}g_{ij}R+\Lambda g_{ij}=2\alpha^2 T_{ij}$,
with $\alpha=0$ corresponding 
to the probe limit discussed in Section 3.
 
  Starting with the dependence of the solutions on the frequency, we have found 
that for a given $\Lambda$, this is fixed by the parameter $\alpha$.
 The spinning solutions with large  enough values of $\alpha$  exhibit the same pattern as in the 
absence of a self-interaction term 
(although with a smaller value of  $ \omega_{min}$),
and the picture in Figure 3 (left) is recovered 
(the same result  was found also for $\Lambda=0$ solutions \cite{Kleihaus:2005me}).
 However, for values of $\alpha$ below a critical value ($i.e.$ for large enough $\lambda$)
the picture for $Q-$balls in an AdS background is recovered. 
  
 \setlength{\unitlength}{1cm}
\begin{picture}(8,6) 
\put(-0.5,0.0){\epsfig{file=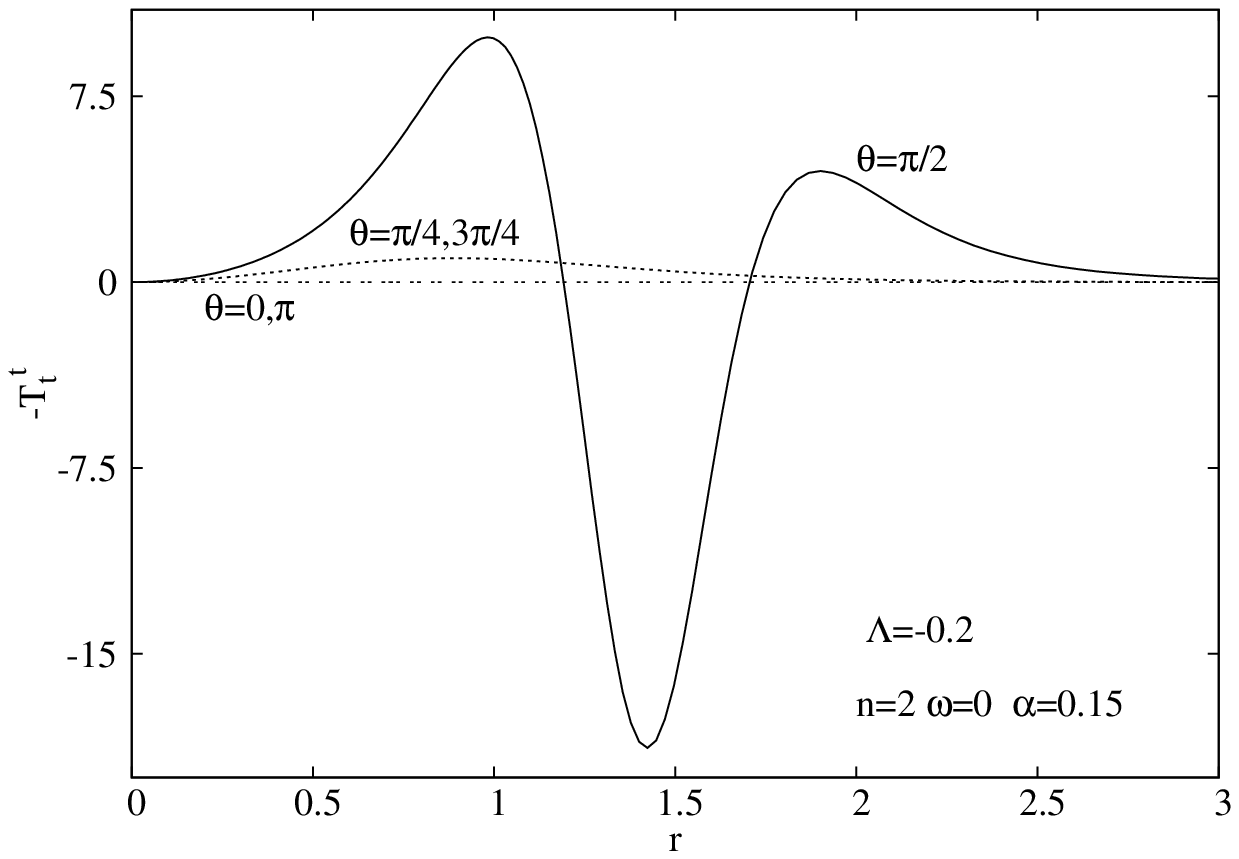,width=8cm}}
\put(8.1,0.0){\epsfig{file=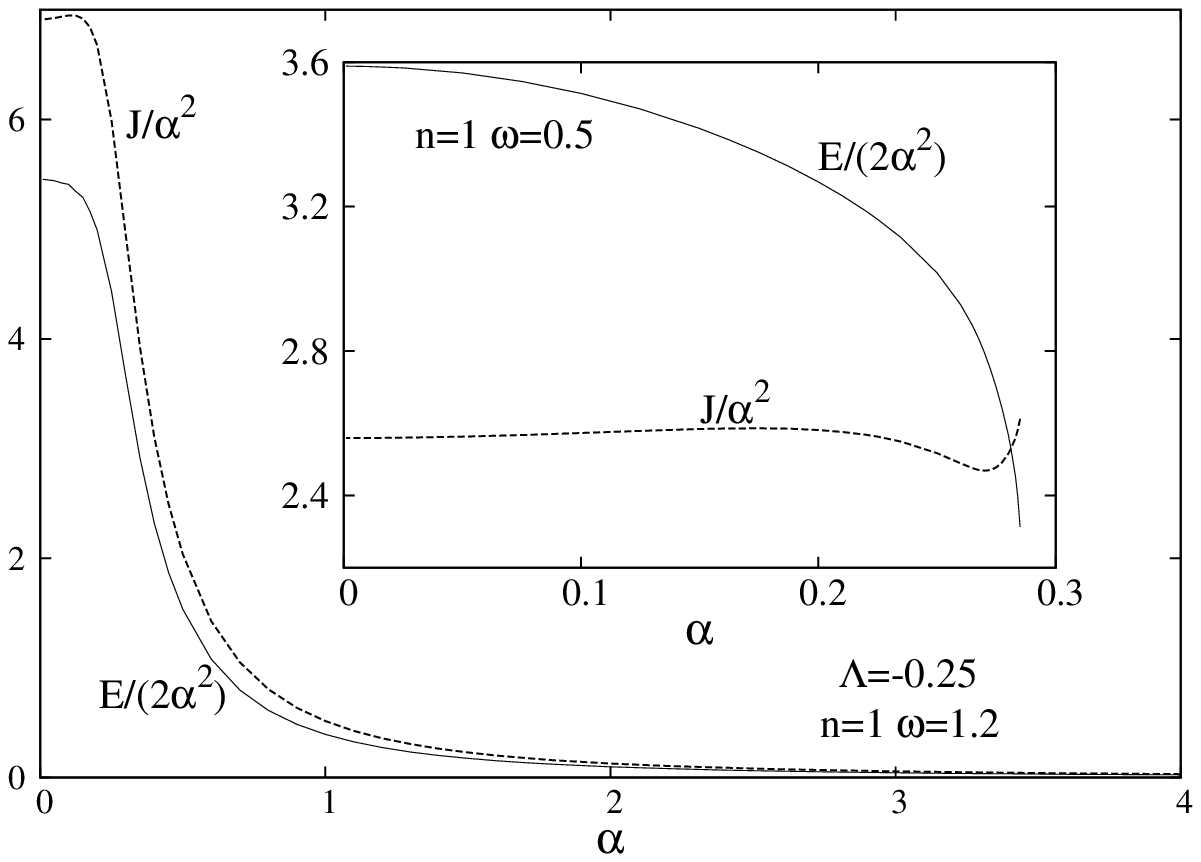,width=8cm}}
\end{picture}
\\
{\small {\bf Figure 4.} 
 {\it Left:}
 The energy density of a static axially symmetric gravitating soliton
 with $ \omega=0$, $n=2$
  is shown for several different angles as a function of the radial coordinate.
{\it Right:} 
 The mass-energy $E$ and the angular momentum $J$ are shown as a function of the  coupling constant
$\alpha=\sqrt{4\pi G \mu^2/\lambda}$
for boson star solutions 
with a $\phi^4$ self interaction potential for two different frequencies.
}
\vspace{0.5cm}
\\
  Thus, for any $n$, the solutions exist for a  range of frequencies $0\leq \omega< \omega_c$, see Figure 3 (right).
In both cases, the configurations with  small $E$, $J$ are just 
perturbations of AdS.
These solitons branch off from the
AdS spacetime for the specific value of the frequency given by (\ref{pert1}).

However, different from the case discussed above with $U(|\Phi|)=\mu^2 |\Phi|^2$,
a self-interaction term in the potential leading to negative energy densities,
allows for a nontrivial limiting solution with $\omega=0$.
This is a new type of soliton, which is different from other solutions with
gravitating scalar fields in the literature 
\cite{Hertog:2004rz}, 
\cite{Hertog:2004ns}, 
\cite{Radu:2005bp}.
Although it is static ($\partial/\partial t$ being a Killing vector of the configuration), 
the geometry has axial symmetry only, being
 regular everywhere, in particular at $r=0$ and on the symmetry axis.
Also, this solution has a 
vanishing Noether charge; however, its mass-energy is finite and nonzero
(see Figure 4 (left) for a plot of the energy density for a typical configuration
with $n=2$; one notices the existence of a region 
with negative energy density, $\rho=-T_t^t<0$).

Concerning the dependence on $\alpha^2=4 \pi G \mu^2/\lambda$, 
a central role is played here by the solutions with a  $|\Phi|^2$ potential only.
We have found that for a range of the frequency  $ \omega_{min}< \omega< \omega_c$ 
(with $ \omega_{min}$ the minimal allowed value
of the frequency for the boson stars without a self-interaction term),
the solutions exist for arbitrarily large values of $\alpha$.
Similar to the asymptotically flat  case \cite{Kleihaus:2005me},
the limit $\alpha \to \infty$
corresponds (after a rescaling) to the solution of the $|\Phi|^2$-model.
The picture is different for smaller frequencies, $ \omega< \omega_{min}$,
in which case there are no solutions in the $|\Phi|^2$ model.
The numerical calculations indicate that, in this case, the range of $\alpha$ is bounded from above,
and a critical configuration is approached for $\alpha \to \alpha_c$,
with $\alpha_c$ depending on $ \omega$ and $\ell$.
This limiting soliton has a finite mass-energy and Noether charge.
(These two cases are illustrated in Figure 4 (right).)
 Unfortunately, 
 the numerical accuracy does not allow  to clarify the  limiting behaviour 
at the critical value of $\alpha$.
We notice only that, as  $\alpha \to \alpha_c$, 
the metric function $F_0$  almost reaches zero at $r=0$,
while the other functions remain finite and nonzero 
(although $F_1$ and $F_2$ take  large values at the origin).

 We remark also that for all spinning solutions,
the distributions of the mass-energy density $-T_t^t$
are very different from those of the spherically symmetric configurations,
$i.e.$ the typical energy density isosurfaces have a toroidal shape.
However, although
 the violation of the positive energy condition is a generic feature of 
 the solutions with $U_{int}=-\lambda|\Phi|^4<0$ (at least for small enough values of $\omega$),
 the mass-energy of all our solutions as given by (\ref{EJ}) is strictly positive.  
 
 We close this Section by noticing that, similar to the $\Lambda=0$ 
 case \cite{Kleihaus:2007vk}, the AdS rotating boson stars possess ergoregions 
in a large part of their domain of existence.
 The ergoregion resides inside the ergosurface  
 defined by the condition $g_{tt}=-F_0N+F_2\sin^2\theta W^2=0$,
 in the metric parametrization (\ref{ansatzg}). 
This type of configurations are typically found for large enough values of $\omega$, $n$ and $\alpha$.
 
\section{Further remarks}
In this work we have initiated a preliminary
investigation of  spinning Q-balls and boson stars in four dimensional AdS spacetime. 
This study was partially motivated by the recent interest in 
solutions of AdS gravity coupled to scalar fields only. 
The picture we have found has some interesting new features as compared to the 
well-known case of solutions with a vanishing cosmological constant.
Perhaps the most interesting new result is the existence of 
axially symmetric solitons with a smooth static limit possessing a vanishing Noether charge.
 Also, all solutions have an upper bound on frequencies,
 which is fixed by the scalar field mass and the cosmological constant.

Moreover,
we expect the existence of a  much richer set of spinning scalar solitons apart from the 
solutions reported in this work.
For example,  the $\Lambda=0$
Q-balls and boson stars with odd parity with respect to 
a reflection in the equatorial plane reported in \cite{Kleihaus:2007vk},
should allow for AdS generalizations.
In particular, it would be interesting 
to construct AdS `{\it twisted}' Q-balls and boson stars,
which combine features of both even and odd parity solutions \cite{Radu:2008pp}.
For such configurations, the scalar field is endowed with a $(r,\theta)$-dependent phase,
$\Phi=\phi (r,\theta)e^{i(n\varphi- \omega t+\Psi(r,\theta))}=(X(r,\theta)+i Y(r,\theta))e^{i(n\varphi- \omega t)}$,
such that the amplitude of the scalar field
vanishes in the equatorial plane.
This would lead to a 'topological charge' of the solutions 
(see \cite{Radu:2008pp} for the details of this
construction in the flat spacetime case).
Also, the issue of AdS vortons, $i.e.$
spinning
vortex loops stabilized by the centrifugal force,
still remains to be investigated.
The results in this work suggest that the  
AdS picture may be very different as compared to 
the one found in the flat spacetime limit \cite{Radu:2008pp}, \cite{Battye:2008mm}.

A natural question which arises concerns the issue 
of higher dimensional counterparts of the solutions discussed in this paper.
Working in the probe limit,
we have found that the general picture we have presented for $d=4$
remains valid 
for spinning solitons 
with a single angular momentum in $d=5,6$ dimensions.
Therefore we expect it to be generic for any $d\geq 4$.
Moreover, for the same self-interaction potential $U_{int}=-\lambda \phi^{2k}$ 
(with $k$ still given by (\ref{k})), the exact Q-ball solution (\ref{ex-sol})
admits a straightforward generalization\footnote{Here we consider a fixed AdS background,
with $ds^2=\frac{dr^2}{1+r^2/\ell^2}+r^2d\Omega_{d-2}^2-(1+r^2/\ell^2)dt^2$.}
 for any $d\geq 3$, with 
$
\Phi(r,t)= (
\frac{\mu^2}{\lambda}\frac{\Delta^2-\ell^2  \omega^2}{(\Delta-(d-1))(\Delta+1))}
 )^{\frac{\Delta}{2}} {\left(1+\frac{r^2}{\ell^2} \right)^{-\frac{\Delta}{2}}} e^{-i  \omega t}, 
$
 and $\Delta=\frac{1}{2}\left( (d-1)\pm \sqrt{(d-1)^2+4\mu^2 \ell^2} \right)$.

Also, based on some preliminary results, 
we conjecture that it is possible to add
a small black hole in the center of the  $d=4$ solitons  with a  harmonic time dependence studied in this work.
Therefore the inclusion of rotation would allow  to circumvent the
no-hair results in \cite{Pena:1997cy}, \cite{Astefanesei:2003qy}.
Indeed,  such solutions 
were constructed recently in \cite{Dias:2011at} for $d=5$
and a complex doublet scalar field,
in which case a special ansatz \cite{Hartmann:2010pm} allows to deal with ODEs.

 
 It would be desirable to study all these solutions
 also
from an AdS/CFT perspective and
to see what they correspond to
in the dual theory.

We close by remarking that
the study of Q-balls and boson stars
is interesting from yet another point of view.
This type of relatively simple 
configurations provide an ideal ground for 
investigating various numerical techniques
on axially symmetric problems with AdS asymptotics,
which thereafter can be applied to more complex models.
\\
\\
\noindent{\textbf{~~~Acknowledgements.--~}} 
We are grateful to Jutta Kunz
for her careful reading of the manuscript and many helpful comments.
 We also thank Burkhard Kleihaus for collaboration in the initial stages of this work. 
We gratefully acknowledge support by the DFG,
in particular, also within the DFG Research
Training Group 1620 ''Models of Gravity''. 
 \begin{small}
 
 \end{small}
 

\begin{thebibliography}{99}
  
\bibitem{Hertog:2004rz}
  T.~Hertog and G.~T.~Horowitz,
  JHEP {\bf 0407} (2004) 073
  [hep-th/0406134].
\bibitem{Hertog:2004ns}
  T.~Hertog and G.~T.~Horowitz,
  Phys.\ Rev.\ Lett.\  {\bf 94} (2005) 221301
  [hep-th/0412169].
\bibitem{Radu:2005bp}
  E.~Radu and E.~Winstanley,
  Phys.\ Rev.\ D {\bf 72} (2005) 024017
  [gr-qc/0503095].
\bibitem{Faulkner:2010fh}
  T.~Faulkner, G.~T.~Horowitz and M.~M.~Roberts,
  Class.\ Quant.\ Grav.\  {\bf 27} (2010) 205007
  [arXiv:1006.2387 [hep-th]].
  
\bibitem{Torii:2001pg}
  T.~Torii, K.~Maeda and M.~Narita,
  Phys.\ Rev.\ D {\bf 64} (2001) 044007.
\bibitem{Sudarsky:2002mk}
  D.~Sudarsky and J.~A.~Gonzalez,
  Phys.\ Rev.\ D {\bf 67} (2003) 024038
  [gr-qc/0207069].
\bibitem{Winstanley:2002jt}
  E.~Winstanley,
  Found.\ Phys.\  {\bf 33} (2003) 111
  [gr-qc/0205092].
\bibitem{Hertog:2004dr}
  T.~Hertog and K.~Maeda,
  JHEP {\bf 0407} (2004) 051
  [hep-th/0404261].
\bibitem{Martinez:2004nb}
  C.~Martinez, R.~Troncoso and J.~Zanelli,
  Phys.\ Rev.\ D {\bf 70} (2004) 084035
  [hep-th/0406111].
  
\bibitem{Park:2008zzb}
  D.~H.~Park,
  Class.\ Quant.\ Grav.\  {\bf 25} (2008) 095002.
  
\bibitem{Pena:1997cy}
  I.~Pena and D.~Sudarsky,
  Class.\ Quant.\ Grav.\  {\bf 14} (1997) 3131.
\bibitem{Kleihaus:2010ep}
  B.~Kleihaus, J.~Kunz, C.~Lammerzahl and M.~List,
  Phys.\ Rev.\ D {\bf 82} (2010) 104050
  [arXiv:1007.1630 [gr-qc]].
\bibitem{Friedberg:1976me}
  R.~Friedberg, T.~D.~Lee and A.~Sirlin,
  Phys.\ Rev.\ D {\bf 13} (1976) 2739.
\bibitem{Coleman:1985ki}
  S.~R.~Coleman,
  Nucl.\ Phys.\ B {\bf 262} (1985) 263
   [Erratum-ibid.\ B {\bf 269} (1986) 744].
\bibitem{Lee:1991ax}
  T.~D.~Lee and Y.~Pang,
  Phys.\ Rept.\  {\bf 221} (1992) 251.
\bibitem{Kusenko:1997zq}
  A.~Kusenko,
  Phys.\ Lett.\ B {\bf 405} (1997) 108
  [hep-ph/9704273].
\bibitem{Kusenko:1997si}
  A.~Kusenko and M.~E.~Shaposhnikov,
  Phys.\ Lett.\ B {\bf 418} (1998) 46
  [hep-ph/9709492].
\bibitem{Jetzer:1991jr}
  P.~Jetzer,
  Phys.\ Rept.\  {\bf 220} (1992) 163;
  \\
  F.~E.~Schunck and E.~W.~Mielke,
  Class.\ Quant.\ Grav.\  {\bf 20} (2003) R301
  [arXiv:0801.0307 [astro-ph]].
\bibitem{Kaup:1968zz}
  D.~J.~Kaup,
  Phys.\ Rev.\  {\bf 172} (1968) 1331.
  
\bibitem{Ruffini:1969qy}
  R.~Ruffini and S.~Bonazzola,
  Phys.\ Rev.\  {\bf 187} (1969) 1767.
\bibitem{Mielke:1980sa}
  E.~W.~Mielke and R.~Scherzer,
  Phys.\ Rev.\ D {\bf 24} (1981) 2111.
\bibitem{Astefanesei:2003qy}
  D.~Astefanesei and E.~Radu,
  Nucl.\ Phys.\ B {\bf 665} (2003) 594
  [gr-qc/0309131].
\bibitem{Prikas:2004yw}
  A.~Prikas,
  Gen.\ Rel.\ Grav.\  {\bf 36} (2004) 1841
  [hep-th/0403019].
\bibitem{Hartmann:2012wa}
  B.~Hartmann and J.~Riedel,
  arXiv:1204.6239 [hep-th].
\bibitem{Horowitz:2010jq}
  G.~T.~Horowitz and B.~Way,
  JHEP {\bf 1011} (2010) 011
  [arXiv:1007.3714 [hep-th]].
\bibitem{Schunck:1996wa}
  F.~E.~Schunck and E.~W.~Mielke,
  Phys.\ Lett.\ A {\bf 249} (1998) 389.
\bibitem{Yoshida:1997qf}
  S.~Yoshida and Y.~Eriguchi,
  Phys.\ Rev.\ D {\bf 56} (1997) 762.
\bibitem{Volkov:2002aj}
  M.~S.~Volkov and E.~Wohnert,
  Phys.\ Rev.\ D {\bf 66} (2002) 085003
  [hep-th/0205157].
\bibitem{Kleihaus:2005me}
  B.~Kleihaus, J.~Kunz and M.~List,
  Phys.\ Rev.\  D {\bf 72} (2005) 064002
  [arXiv:gr-qc/0505143].
\bibitem{Kleihaus:2007vk}
  B.~Kleihaus, J.~Kunz, M.~List and I.~Schaffer,
  Phys.\ Rev.\ D {\bf 77} (2008) 064025
  [arXiv:0712.3742 [gr-qc]].
\bibitem{Astefanesei:2003rw}
  D.~Astefanesei and E.~Radu,
  Phys.\ Lett.\ B {\bf 587} (2004) 7
  [gr-qc/0310135].
\bibitem{Dias:2011at}
  O.~J.~C.~Dias, G.~T.~Horowitz and J.~E.~Santos,
  JHEP {\bf 1107} (2011) 115
  [arXiv:1105.4167 [hep-th]].
\bibitem{Hartmann:2010pm}
  B.~Hartmann, B.~Kleihaus, J.~Kunz and M.~List,
  Phys.\ Rev.\ D {\bf 82} (2010) 084022
  [arXiv:1008.3137 [gr-qc]].
\bibitem{Colpi:1986ye}
  M.~Colpi, S.~L.~Shapiro and I.~Wasserman,
  Phys.\ Rev.\ Lett.\  {\bf 57} (1986) 2485.
  
\bibitem{Henneaux:2006hk}
  M.~Henneaux, C.~Martinez, R.~Troncoso and J.~Zanelli,
  Annals Phys.\  {\bf 322} (2007) 824
  [hep-th/0603185].
\bibitem{Henneaux:2004zi}
  M.~Henneaux, C.~Martinez, R.~Troncoso and J.~Zanelli,
  Phys.\ Rev.\ D {\bf 70} (2004) 044034
  [hep-th/0404236].
\bibitem{Breitenlohner:1982jf}
  P.~Breitenlohner and D.~Z.~Freedman,
  Annals Phys.\  {\bf 144} (1982) 249;
\\
  P.~Breitenlohner and D.~Z.~Freedman,
  Phys.\ Lett.\ B {\bf 115} (1982) 197.
\bibitem{Balasubramanian:1999re}
V.~Balasubramanian and P.~Kraus,
Commun.\ Math.\ Phys.\  {\bf 208} (1999) 413
[arXiv:hep-th/9902121].
\bibitem{Ash}
  A.~Ashtekar and A.~Magnon,
  Class.\ Quant.\ Grav.\  {\bf 1} (1984) L39;
\\
  A.~Ashtekar and S.~Das,
  Class.\ Quant.\ Grav.\  {\bf 17} (2000) L17
  [arXiv:hep-th/9911230].
  
\bibitem{Kleihaus:1996vi}
  B.~Kleihaus and J.~Kunz,
  Phys.\ Rev.\ Lett.\  {\bf 78} (1997) 2527
  [hep-th/9612101];
\\
  B.~Kleihaus and J.~Kunz,
  Phys.\ Rev.\ Lett.\  {\bf 79} (1997) 1595
  [gr-qc/9704060];
  \\
  B.~Kleihaus and J.~Kunz,
  Phys.\ Rev.\ Lett.\  {\bf 86} (2001) 3704
  [gr-qc/0012081].
\bibitem{Wiseman:2002zc}
  T.~Wiseman,
  Class.\ Quant.\ Grav.\  {\bf 20} (2003) 1137
  [arXiv:hep-th/0209051].
\bibitem{schoen}
 W. Sch\"onauer and R. Wei\ss ,
 J. Comput. Appl. Math. 27, 279 (1989) 279;
 \\
 M. Schauder, R. Wei\ss\ and W. Sch\"onauer,
 The CADSOL Program Package,
 Universit\"at Karlsruhe, Interner Bericht Nr. 46/92 (1992).   
     
\bibitem{Radu:2008pp}
  E.~Radu and M.~S.~Volkov,
  Phys.\ Rept.\  {\bf 468} (2008) 101
  [arXiv:0804.1357 [hep-th]].
   
\bibitem{Kusenko:1997ad}
  A.~Kusenko,
  Phys.\ Lett.\ B {\bf 404} (1997) 285
  [hep-th/9704073].
\bibitem{Tamaki:2010zz}
  T.~Tamaki and N.~Sakai,
  Phys.\ Rev.\ D {\bf 81} (2010) 124041
  [arXiv:1105.1498 [gr-qc]].
\bibitem{Battye:2008mm}
  R.~A.~Battye and P.~M.~Sutcliffe,
  Nucl.\ Phys.\ B {\bf 814} (2009) 180
  [arXiv:0812.3239 [hep-th]].

 


















        

 \end{thebibliography}
 \end{document}